\documentclass[twocolumn,prb,showpacs,preprintnumbers,superscriptaddress,amsmath,amssymb]{revtex4}

\usepackage{color}
\usepackage{graphicx}

\begin{document}


\title{Heterogeneous spin state in the field-induced phase of volborthite as seen
via $^{51}$V nuclear magnetic resonance}


\author{M. Yoshida}
\affiliation{Institute for Solid State Physics, University of Tokyo, Kashiwa, Chiba 277-8581, Japan}

\author{M. Takigawa}
\affiliation{Institute for Solid State Physics, University of Tokyo, Kashiwa, Chiba 277-8581, Japan}
\author{H. Yoshida}
\affiliation{Superconducting Materials Center, National Institute for Materials Science (NIMS), Tsukuba, Ibaraki 305-0044, Japan}
\author{Y. Okamoto}
\affiliation{Institute for Solid State Physics, University of Tokyo, Kashiwa, Chiba 277-8581, Japan}
\author{Z. Hiroi}
\affiliation{Institute for Solid State Physics, University of Tokyo, Kashiwa, Chiba 277-8581, Japan}


\date{\today}

\begin{abstract}
We report results of $^{51}$V NMR in the field-induced phase of volborthite 
Cu$_3$V$_2$O$_7$(OH)$_2 \cdot $2H$_2$O, a spin-1/2 antiferromagnet 
on a distorted kagome lattice. In magnetic fields above 4.5~T, two types of 
V sites with different spin-echo decay rates are observed. The hyperfine field at 
the fast decaying sites has a distribution, while it is more homogeneous at the 
slowly decaying sites. Our results indicate a heterogeneous state consisting of 
two spatially alternating Cu spin systems, one of which exhibits anomalous spin fluctuations 
contrasting with the other showing a conventional static order.
\end{abstract}


\maketitle



Possibility of exotic quantum states in two-dimensional (2D) spin systems 
with frustrated interactions have attracted strong attention.\cite{Misguich,IFM} 
In particular, the ground state of the spin-1/2 Heisenberg model 
with a nearest-neighbor antiferromagnetic (AF) interaction on a kagome 
lattice, a 2D network of corner-sharing equilateral triangles, is believed 
to develop no long-range magnetic order. Theories have proposed various ground 
states such as spin liquids with no broken symmetry with \cite{Waldtmann} or 
without \cite{Hermele} an excitation gap or symmetry breaking valence-bond-crystal 
states.\cite{Singh} Candidate materials known to date, however, depart 
from the ideal kagome model in some aspects, such as disorder, structural distortion, 
anisotropy, or longer range interactions. Effects of the Dzyaloshinsky-Moriya 
(DM) interaction,\cite{Cepas} spatially anisotropic exchange interactions,\cite{Schnyder,Wang,Stoudenmire} 
and longer range Heisenberg interactions \cite{Domenge} have been theoretically investigated. 

Volborthite Cu$_3$V$_2$O$_7$(OH)$_2 \cdot $2H$_2$O is a layered compound, in which 
distorted kagome layers are formed by isosceles triangles of Cu ions carrying a spin 1/2. 
Consequently, there are two Cu sites, Cu1 and Cu2, and two kinds of exchange 
interactions, $J$ and $J'$, as shown in Fig.~1. The Cu2 sites form linear chains, which 
are connected through the Cu1 sites. The magnetic susceptibility $\chi $ obeys the Curie-Weiss 
law $\chi  = C/(T + \theta _W)$ above 200~K with $\theta _W$ = 115 K, exhibits a broad 
maximum at 20~K, and approaches a finite value at the lowest temperatures.\cite{Hiroi,HYoshida} 
An unusual magnetic transition has been observed near 1~K,\cite{Fukaya,Bert,MYoshida,Yamashita} 
which is much lower than $\theta _W$, consistent with strong frustration in a 
kagome lattice.  However, a recent density functional calculation \cite{Janson} 
proposed a ferromagnetic $J$ and a sizable AF interaction $J''$ between second 
neighbors along the chain as shown in Fig.~1. In this model, frustration arises from 
the competition between $J$ and $J''$ along the chain rather than the geometry of a kagome 
lattice. The appropriate spin model for volborthite has not been settled yet. 
Recently, anomalous sequential magnetization 
steps were reported in a high quality polycrystalline sample at 4.3, 25.5, and 46~T.\cite{HYoshida} 
Kaneko \textit{et al}. showed a classical Heisenberg model on a spatially distorted kagome lattice 
exhibits a magnetization step by the Monte Carlo method.\cite{Kaneko} 

\begin{figure}[b]
\includegraphics[width=0.7\linewidth]{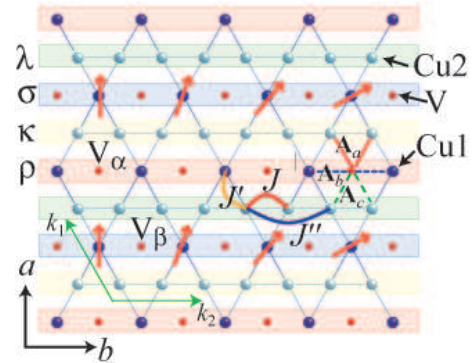}
\caption{\label{fig:fig1} (color online) Schematic structure of volborthite 
projected onto the $a$-$b$ plane. The H and O sites are not shown. 
The V sites are located below and above the Cu kagome layers. 
The upper and lower V sites are related by inversion with respect to the Cu sites, 
hence all V sites are equivalent. See the text for the definition of various symbols.
}
\end{figure}

A magnetic transition in volborthite at $T^* \sim 1$~K was detected by 
$^{51}$V-NMR,\cite{Bert,MYoshida} muon spin relaxation,\cite{Fukaya} 
and heat capacity \cite{Yamashita} experiments. Our previous NMR 
measurements \cite{MYoshida} revealed a sharp peak in the nuclear 
spin-lattice relaxation rate $1/T_1$ and onset of broadening of the 
NMR spectrum due to development of spontaneous moments 
for the magnetic field $B$ below 4.5~T.  A kink was also observed 
in the heat capacity.\cite{Yamashita} Most recently, development of short 
range spin correlation was detected by neutron inelastic scattering.\cite{Nilsen} 
However, the low $T$ phase, which we call phase I, shows various anomalies incompatible with a 
conventional magnetic order.\cite{MYoshida,Yamashita} The NMR line shape 
is not rectangular but fits to a Lorentzian (Fig.~2b), suggesting spatial 
modulation of static moments. The behavior $1/T_1 \propto T$ provides 
evidence for dense low-energy excitations. 
The spin-echo decay rate $1/T_2$ is anomalously large, pointing to unusually slow 
spin fluctuations. 

Above 4.5 T, at which the first magnetization step occurs, another magnetic phase 
appears with the different NMR line shape and dynamics.\cite{MYoshida} 
The spin state in this phase, which we call phase II, has not been well characterized yet. 
In this paper, we report results of $^{51}$V NMR in phase II, which revealed 
coexistence of two types of V sites with different $1/T_2$ and line shapes. 
Our results indicate that phase II exhibits a unique heterogeneous spin state, where 
two distinct types of Cu spins, one with non-uniform magnitude and large temporal 
fluctuations and the other with a more conventional static order, form a periodic structure. 

$^{51}$V NMR measurements have been performed on a high-quality powder 
sample synthesized as described in Ref.~\onlinecite{HYoshida}. The NMR spectra were 
obtained by summing the Fourier transform of the spin-echo signal 
at equally spaced frequencies in a fixed $B$. 
The pulse sequence $\pi /2 - \tau  - \pi /2$ was used with 
the pulse width of 1-3 $\mu$s and the pulse separation $\tau $ in the range of 10-150 $\mu$s. 
We determined $1/T_2$ by fitting the spin-echo intensity $I(\tau )$ as a function of $\tau$ 
to an exponential decay $I(\tau ) = A_0\mathrm{exp}(-2\tau /T_2)$. 
Likewise, $1/T_1$ was obtained by fitting the nuclear recovery curve $I(t)$, 
which is the spin-echo intensity as a function of the time $t$ between the 
saturating comb pulses and the first $\pi/2$ pulse, to the exponential function 
$I(t) = I_{eq} - I_0\mathrm{exp}(-t/T_1)$.  When this function did not fit the 
data due to inhomogeneous distribution of $1/T_1$, we used the stretched exponential 
function $I(t) = I_{eq} - I_0\mathrm{exp}\{-(t/T_1)^{\beta }\}$ to determine 
the representative value of $1/T_1$. 

Figure~2(a) shows the $\tau $ dependence of the NMR spectra at 6.0~T 
and 0.3~K (phase II). The spectrum for $\tau $ = 10 $\mu$s is identical to 
one of the spectra shown in Fig.~4(a) of Ref.~\onlinecite{MYoshida} except that 
it was plotted against the shift in resonance field rather than the frequency 
in Ref.~\onlinecite{MYoshida}.  In addition to the overall decay of the 
intensity with $\tau$, the line shape also changes with $\tau$. 
The broad peak near the center of the spectrum 
observed for $\tau $ = 10 and 25 $\mu$s disappears for $\tau \ge  40$ $\mu$s. 
Instead the two peaks indicated by the arrows become more pronounced with 
increasing $\tau $. Such $\tau$ dependence indicates that the spectrum 
consists of two or more components characterized by different values of $1/T_2$. 
On the other hand, the line shape at 1~T (phase I) does not depend on $\tau$ 
as shown in Fig.~2(b). 

\begin{figure}
\includegraphics[width=0.95\linewidth]{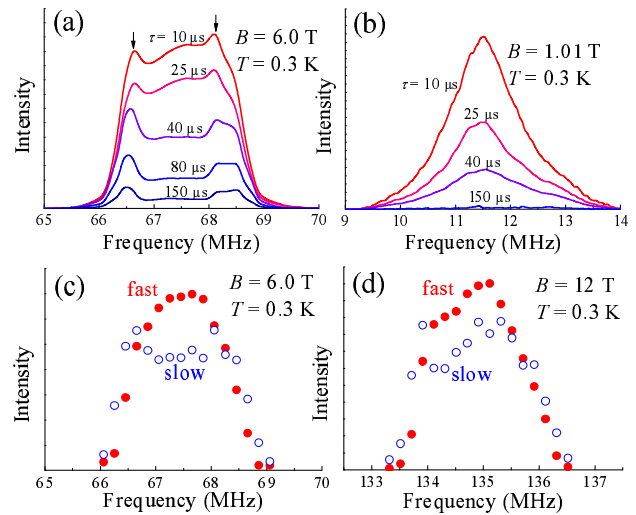}
\caption{\label{fig:fig2} (color online) NMR spectra for various values of $\tau $ at 0.3~K 
in the magnetic field of (a) 6 T and (b) 1.01 T. The spectra in (c) and (d) show 
decomposition of the NMR spectra at 0.3 K into the fast decaying (solid circles) and 
slowly decaying (open circles) components for the magnetic field of 6~T and 12~T.}
\end{figure}

We confirmed the two-component behavior in phase II by direct measurements of 
$1/T_2$.  Figure~3 shows the spin-echo decay curves at 0.3~K measured at the 
center of the spectrum obtained in the field of 1, 6, and 12~T. 
The data at 1 T can be fit to the single exponential function, providing 
unique $1/T_2$. On the other hand, the data at 6 and 12~T show a convex 
curve, which can be well fit to the two-component exponential function 
\begin{eqnarray}
I(\tau ) = A_f\mathrm{exp}(-2\tau /T_{2f}) + A_s\mathrm{exp}(-2\tau /T_{2s}) 
\end{eqnarray}
as shown by the solid lines. 
The fast and slowly decaying components at 6~T are shown 
separately by the dashed lines.  The decay rate of the fast component, $1/T_{2f}$, 
is nearly equal to $1/T_2$ at 1~T. 

We define the two V sites showing the fast and slow decay rates as the V$_f$ and 
V$_s$ sites, respectively. The coefficients $A_f$ and $A_s$ in eq.~(1) are proportional 
to the number of these sites within the frequency window covered 
by the exciting rf-pulse.  Therefore, the decomposition 
of a NMR spectrum into the fast and slow components can be accomplished 
by measuring the spin-echo decay curve at many frequency points covering 
the entire spectrum and plotting $A_f$ and $A_s$ against frequency. 
The decomposed spectra thus obtained at 6 and 12~T are shown in Fig.~2(c) and (d). 
The pulse width was set to 3~$\mu$s covering about $\pm$0.1~MHz. The 
two-component fit of the spin-echo decay curve was successful at all 
frequencies. The integrated intensities of the ``fast'' and ``slow'' spectra correspond 
to the total number of the V$_f$ and V$_s$ sites.  For the spectra shown in 
Fig.~2(c) and (d), the integrated intensities of the two components are nearly equal 
within the experimental error, indicating equal abundance of the V$_f$ and V$_s$ sites. 

\begin{figure}
\includegraphics[width=0.8\linewidth]{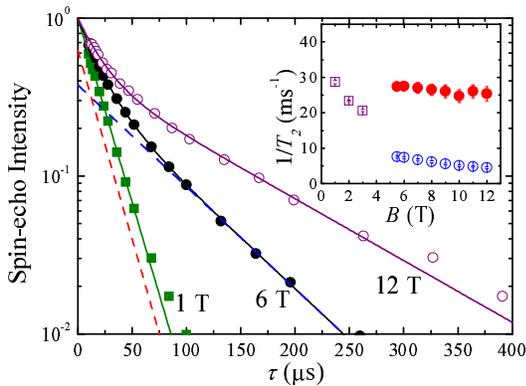}
\caption{\label{fig:fig3} (color online) Spin-echo decay curves at 0.3 K 
in the magnetic field of 1, 6, and 12~T.  The fast and slowly decaying components 
for the 6~T data are shown by the dashed lines.  The inset shows the $B$ 
dependence of $1/T_2$ at 0.3 K.  The open squares indicate $1/T_2$ in phase I. 
The solid and open circles represent $1/T_{2f}$ and $1/T_{2s}$ in phase II.}
\end{figure}

The inset of Fig.~3 shows the $B$ dependence of $1/T_2$ at 0.3 K 
measured at the center of the spectrum. In phase I, 
the spin-echo decay curve can be always fit to the single exponential 
function. Such a clean behavior is brought by the improvement of sample 
quality,\cite{HYoshida} since two-component behavior was observed even 
near 2~T in earlier studies.\cite{Bert} The decrease 
of $1/T_2$ with increasing field indicates suppression of slow spin 
fluctuations by magnetic fields. On the other hand, $1/T_1$ is insensitive 
to $B$ in phase I.\cite{MYoshida}
In phase II, the decay curve can be well fit to the 
two-component function (1). Both $1/T_{2f}$ and $1/T_{2s}$ slightly decreases 
with increasing $B$. Near the transition at 4.5~T, the two-component fit 
does not work well probably due to coexistence of the two phases. 

Figure~4(a) shows the $T$ dependence of 1/$T_2$ at 6 T together with the 
previous data at 1~T.\cite{MYoshida} The spin-echo decay at 1~T follows 
the single exponential function at all temperatures (crosses). 
At 6~T, the decay curve can be fit by the single exponential function for $1.5 \le T \le 2$~K. 
When 1/$T_2$ becomes small for $T >$ 2.0~K, the Gaussian contribution to 
the spin-echo decay due to the nuclear spin-spin coupling has to be taken into account.  
Then the decay curve can be fit to the ``exponential~+~Gaussian'' function $I(\tau ) = A_0\mathrm{exp}\{-2\tau /T_2-2(\tau /T_{2G})^2\}$. The Gaussian decay rate is almost temperature independent 
($1/T_{2G}$ = 2 $\times 10^3$ s$^{-1}$). 
The decay curve exhibits the two-component exponential behavior below 1.5~K, 
which is the boundary between the uniform paramagnetic phase and phase II. 
We then determined $1/T_{2f}$ (solid circles) and $1/T_{2s}$ (open circles) by 
using eq.~(1). Although  $1/T_{2f}$ and $1/T_{2s}$ are different 
in magnitude, they show similar $T$ dependences and decrease slightly 
with decreasing $T$. This $T$-dependence is nearly identical to the behavior 
at 1~T (Fig.~4a). Below 1.1 K, the ratio $A_s/A_f$ is almost $T$-independent, 
indicating that the numbers of the V$_f$ and V$_s$ sites stays nearly equal. 
The two-component fit does not work well near 1.5~K probably due to 
coexistence of the paramagnetic phase and phase II. 

Figure~4(b) shows the $T$ dependence of $1/T_1$ at 6~T (triangles) together with 
the previous data at 1~T (asterisks).\cite{MYoshida} The recovery curve at 1~T follows 
the exponential function for $T \ge 1$~K, while the stretched exponential function 
is required to fit the data at lower temperatures.\cite{MYoshida} Although the 
recovery curves do not depend on $\tau$ at 1~T (phase I), they do in phase II. 
The open and solid triangles represent $1/T_1$ at 6~T for $\tau $ = 10 and 
80~$\mu$s, respectively. Above 1.5~K, the recovery curve can 
be reproduced by the exponential function and $1/T_1$ does not depend on $\tau $. 
Below 1.5~K (phase II) the recovery curve for $\tau $ = 10~$\mu$s 
was fit to the stretched exponential function. We then obtained 
$1/T_1 \propto T^{1.5}$ behavior,\cite{MYoshida} but $\beta $ depends on $T$ and 
reaches an extremely small value of 0.25 at 0.14 K. 
This can be now accounted for by coexistence of the V$_f$ and V$_s$ sites for short $\tau$. 
The recovery curve for $\tau $ = 80~$\mu$s can be fit to the stretched exponential 
function with constant $\beta  \approx  0.6$, which is also the case at 1 T. 
Since the signal from the V$_f$ sites has decayed at long $\tau$ and 
is nearly absent at $\tau $ = 80~$\mu$s, 
$1/T_1$ thus obtained should correspond to the V$_s$ sites. 
The solid line in Fig.~4(b) shows a $1/T_1 \propto T$ behavior, which fits 
the data for both 6~T and 1~T. This indicates that the V$_s$ sites in phase II are 
affected by anomalously dense low-energy excitations that have been identified 
in phase I in Ref.~\onlinecite{MYoshida}. The smaller values of $1/T_1$ at 6~T 
indicates that the V$_s$ sites couple to these excitations more weakly than the V 
sites in phase I. 

\begin{figure}
\includegraphics[width=0.8\linewidth]{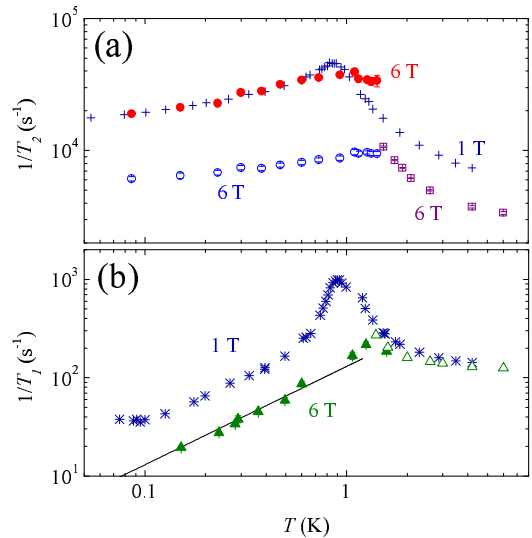}
\caption{\label{fig:fig4} (color online) The $T$ dependences of (a) $1/T_2$ 
and (b) $1/T_1$. The various symbols are defined in the text. 
Two components of $1/T_2$ appear at 6~T below 1.5~K, whereas only 
the slow component can be determined for $1/T_1$. The data at 1~T are taken from 
Ref.~\onlinecite{MYoshida}.}
\end{figure}

How can we understand the spin state in phase II? The V$_f$ and V$_s$ sites show 
significantly different relaxation rates and line shapes. 
The spectra at the V$_f$ sites show a round shape (Fig.~2cd), 
indicating a distribution in  the magnitude of the hyperfine 
field, which can be generated, for instance, by a spin-density-wave 
order or spatially disordered static moments. Therefore, the large 
$1/T_{2f}$ at the V$_f$ sites should be ascribed to unusually slow fluctuations 
of the Cu spins with non-uniform moments. 
On the other hand, the V$_s$ sites show a rectangular spectral
shape, which is compatible with conventional AF order. 
The relatively small $1/T_{2s}$ and $1/T_1$ show that the effects of the unusual spin fluctuations 
are weak at the V$_s$ sites. 

How are the two V sites distributed spatially? Two cases can be considered. 
In one case, a magnetic superstructure is formed in phase II yielding two inequivalent 
V sites. Alternatively, macroscopic phase separation may occur and each phase 
contains only one of the two V sites. Our NMR results definitely support the first case. 
The numbers of the V$_f$ and V$_s$ sites are nearly equal in the entire region 
of phase II, which covers a wide range of $B$ and $T$. This is quite natural 
if a microscopic superstructure is formed but very unlikely for the phase 
separation scenario, since the relative stability of the two phases should change 
with $B$ and $T$. In addition, the two V sites show similar $T$ dependences 
of $1/T_2$ in spite of the difference in magnitude (Fig.~4a). 
This indicates a common source of the fluctuating hyperfine field for the 
two V sites. The different magnitude should then be attributed to the 
difference in the hyperfine coupling between the fluctuating spins and the nuclei. 
Since the range of hyperfine interaction is short with an atomic scale, 
this is not possible for the case of phase separation. 

Although we cannot determine the superstructure from the NMR data on 
a powder sample alone, let us discuss most likely possibilities. 
Since the numbers of the two V sites are equal, we expect a doubled unit cell 
due to formation of a superstructure.  For the ideal kagome 
lattice, the three directions ${\bf k}_1$, ${\bf k}_2$, and ${\bf k}_1 + {\bf k}_2$ 
(Fig.~1) are all equivalent. However, since the structure of volborthite 
is uniform along ${\bf k}_2$ and inequivalent Cu1 and Cu2 sites 
alternate along ${\bf k}_1$, we consider the wave vector of the superstructure 
is likely to be along ${\bf k}_1$.  The two V sites then alternate along 
${\bf k}_1$. The V$_{\alpha }$ and V$_{\beta }$ in Fig.~1 correspond 
to the V$_f$ and V$_s$ sites.   Likewise, each of the Cu1 and Cu2 sites 
are divided into two inequivalent sites Cu1$_{\rho}$, Cu1$_{\sigma}$ 
and Cu2$_{\kappa}$, Cu2$_{\lambda }$, as shown in Fig. 1. 

Since a V site is located approximately above or below the center of the 
Cu hexagon of the kagome lattice, the dominant source of the hyperfine field 
at V nuclei should be confined within the six Cu spins on a hexagon. Because 
of the distorted structure of volborthite, there are three distinct hyperfine 
coupling tensors ${\bf A}_a$, ${\bf A}_b$, and ${\bf A}_c$ as shown in Fig.~1. 
The coupling tensors to the other three spins ${\bf A}_a'$, ${\bf A}_b'$, and 
${\bf A}_c'$ are generated by the mirror reflection perpendicular to the $b$ 
axis at the V site. The hyperfine fields at V$_{\alpha }$ and V$_{\beta }$ 
are written as 
${\bf B}_{\alpha } = ({\bf A}_a{\bf s}_{\kappa } + {\bf A}_a'{\bf s}_{\kappa }') 
+ ({\bf A}_b{\bf s}_{\rho } + {\bf A}_b'{\bf s}_{\rho }') 
+ ({\bf A}_c{\bf s}_{\lambda } + {\bf A}_c'{\bf s}_{\lambda }') $
and ${\bf B}_{\beta } = ({\bf A}_a{\bf s}_{\lambda } + {\bf A}_a'{\bf s}_{\lambda }') 
+ ({\bf A}_b{\bf s}_{\sigma } + {\bf A}_b'{\bf s}_{\sigma }') 
+ ({\bf A}_c{\bf s}_{\kappa } + {\bf A}_c'{\bf s}_{\kappa }') $, 
respectively, where ${\bf s}_{\epsilon }$ and ${\bf s}_{\epsilon }'$ denote 
two neighboring spins on the same type of sites. ($\epsilon$ stands for Cu1$_{\rho}$, 
Cu1$_{\sigma}$, Cu2$_{\kappa}$, or Cu2$_{\lambda }$.) 

Both $1/T_1$ and $1/T_2$ are determined by the time-correlation function 
of the hyperfine field.\cite{Slichter} If all the hyperfine coupling tensors ${\bf A}_a$, 
${\bf A}_b$, and ${\bf A}_c$ are largely isotropic and have similar magnitudes, the 
different $1/T_2$ for the two V sites must be ascribed to the difference in the 
time-correlation functions of ${\bf s}_{\rho }$ and ${\bf s}_{\sigma }$. A plausible 
situation is illustrated in Fig.~1. The Cu1$_{\sigma }$ sites develop a 
conventional long range magnetic order. Hence, the spin fluctuations of these sites 
have a small amplitude with a conventional time scale typically described by 
spin waves. On the other hand, the Cu1$_{\rho }$ sites show modulated spin 
structure. The absence of fully developed static moments would allow unusually 
slow fluctuations responsible for the large $1/T_2$ at the V$_f$ sites. 
By studying the anisotropic kagome model in the limit $J \gg  J^{\prime}$, 
Schnyder \textit{et al.} proposed a spiral order with large (small) ordered moments on the Cu1 
(Cu2) sites.\cite{Schnyder} Although the different magnitudes of the ordered moments  
for the two Cu sites is a feature in common to our results, our observation of two 
distinct V sites requires an enlarged magnetic unit cell. 
The spin structure of the Cu1$_{\sigma}$ sites could be a spiral 
order with a small wave vector as proposed theoretically.\cite{Stoudenmire,Schnyder} 

Alternatively, the low symmetry of the volborthite structure may result in large 
difference in the hyperfine coupling. In particular, ${\bf A}_a$ and ${\bf A}_c$ 
involve significantly different V-O-Cu hybridization paths. If one of them, say ${\bf A}_a$, 
is the dominant coupling, the different $1/T_2$ for the two V sites must be ascribed 
to the different dynamics of ${\bf s}_{\lambda }$ and ${\bf s}_{\kappa }$ spins. 

In conclusion, our NMR results show coexistence of the two types of 
V sites with different relaxation behavior and spectral shapes 
in the field-induced phase II of volborthite. Our results indicate 
that the high field phase of volborthite exhibits a unique heterogeneous 
magnetic state, where a non-uniform order with anomalous fluctuations 
alternates with a more conventional static order. 

We thank M. Horvati\'{c} and C. Berthier for stimulating discussions. 
The work was supported by JSPS KAKENHI B (21340093), MEXT KAKENHI on 
Priority Areas ``Novel States of Matter Induced by Frustration'' (22014004) and 
by the MEXT-GCOE program.

\end{document}